\documentclass{ws-procs9x6-cpt25}
\begin{document}


\newcommand{\refeq}[1]{(\ref{#1})}
\def\etal {{\it et al.}}

\title{New limits on CPT-Symmetry Violation \\
in Charm mesons}

\author{W. Krzemie\'n$^1$,M .Kmieć$^{2}$, A. Szabelski$^{1}$ and W.\ Wi\'slicki$^2$}
\address{$^1$High Energy Physics Division, National Centre for Nuclear Research,\\
Andrzeja Soltana 7, Otwock, Swierk, PL-05-400, Poland}
\address{$^2$Department of Complex Systems, National Centre for Nuclear Research,\\
Andrzeja Soltana 7, Otwock, Swierk, PL-05-400, Poland}



\begin{abstract}
Neutral flavour meson oscillations are a fascinating quantum phenomenon that allows for high-precision tests of fundamental symmetries, including CPT invariance.
In this contribution, we present the methodology for testing CPT violation in neutral flavour meson systems, with particular emphasis on the charm sector. We report the recently established 
constraints on CPT-violating effects, derived from a reinterpretation of the LHCb measurement of the time-dependent asymmetry in the Cabibbo-favoured decays $D^0 \rightarrow K^{-}\pi^{+}$ and $\overline{D^{0}} \rightarrow K^{+}\pi^{-}$.   Our results improve the previous bounds by the FOCUS experiment by approximately two orders of magnitude.
We also discuss prospects for tests of CPT invariance, including the Standard Model Extension framework. 
\end{abstract}

\bodymatter

\section{Introduction}

Neutral flavour meson oscillations represent a remarkable quantum interference phenomenon, offering a sensitive probe to test fundamental symmetries such as the CPT invariance. Among known systems, neutral mesons, especially kaons, enable experimental sensitivity approaching the Planck scale~\cite{particledatagroupReviewParticlePhysics2022}.

The time evolution of neutral mesons is typically described using the Weisskopf-Wigner approximation~\cite{weisskopfBerechnungNaturlichenLinienbreite1930}, where the dynamics is governed by the 2×2 non-Hermitian effective Hamiltonian. CPT violation can be parametrised via a complex parameter $z$. It can be expressed (see \emph{e.g.} Ref.~\refcite{kosteleckyCPTLorentzViolation2001}) as the ratio between the difference of the diagonal elements of the effective Hamiltonian, encoding mass ($\delta m$) and decay width ($\delta \Gamma$) differences between the flavour states, and the normalised differences of the mass ($x=\frac{\Delta M}{\Gamma})$ and decay width ($y = \frac{\Delta \Gamma}{2\Gamma}$) eigenvalues of the meson mass states. The $\Gamma$ parameter is the average decay rate.

\begin{equation}
\label{zpq}
 z = \frac{\delta m - i \delta \Gamma/ 2 }{\Gamma(x-iy)}.
\end{equation}
The parameter $z$ is non-zero if and only if CPT is violated. (see Refs.~\refcite{krzemienNewConstraintsSymmetry2024,amhisAveragesHadronHadron2023}).  $x, y$ are often called mixing parameters because they govern the oscillation behaviour of the neutral meson system.

In the classical approach, the parameter $z$ is treated phenomenologically, without an underlying theoretical interpretation. Alternatively, $z$ can be derived within a theoretical framework such as the Standard-Model Extension (SME)~\cite{colladayCPTViolationStandard1997}, which connects it to fundamental CPT- and Lorentz-violating mechanisms.
Although the same formalism is used to describe all four neutral meson families ($K^{0}, D^{0}, B^{0}, B_{s}^{0}$), their phenomenology varies significantly. This is due to the differences in oscillation frequencies and decay widths across the meson families. For example, the $B_s$ meson changes its flavour approximately 26 times within its lifetime, while oscillations in the charm sector are extremely slow, with the frequency of the order of 0.1\%.

Various experimental strategies exist to study CPTV in these systems, depending on whether the mesons are produced in correlated pairs (originating from quarkonium decays e.g. $\phi(1020)\rightarrow K^{0} \bar{K}^{0}$, $\psi(3770) \rightarrow D^{0} \bar{D}^{0}$, $\Upsilon(4S) \rightarrow B^{0} \bar{B}^{0}$ ) or as uncorrelated particles, and whether they decay into flavour-specific or common final states~\cite{Bernabeu:2003ym, robertsTestingSymmetryCorrelated2017,vantilburgStatusProspectsCPT2015,kosteleckyCPTLorentzViolation2001}. 

\section{CPT in charm sector}

The charm sector offers a distinct and complementary environment for CPTV studies compared to the strange and beauty sectors. The mass of the charm quark ($\approx 1.27$ GeV/c~$^2$) places it in the intermediate regime between light ($u$, $d$, $s$) and heavy ($b$, $t$) quarks, where non-perturbative QCD effects play a significant role. This fact, combined with various suppression mechanisms, makes theoretical predictions in charm particularly challenging \cite{fridayCharmPhysics2025}.
In the Standard Model, both quantum oscillations and CP-violating effects in the charm sector are expected to be highly suppressed, typically of the order $10^{-3}-10^{-4}$, see Ref.~\refcite{lenzMixingCPViolation2021}. 

From an experimental point of view, several features specific to the charm sector complicate the CPTV studies. Due to the short lifetime of $D^{0}$ mesons ($\tau_{D^{0}} = 410$~fs ) and slow oscillation rates, experiments must collect large datasets while maintaining high signal purity. This requires efficient and selective trigger systems, precise vertex reconstruction to distinguish prompt mesons from secondary ones (originating from B decays), excellent momentum resolution, and robust particle identification.
Time-dependent analyses additionally require high Lorentz particle boosts and detectors with very high time resolutions.
These conditions are exceptionally well fulfilled by the LHCb experiment, which provides large charm datasets and decay time resolutions at the $0.1 \tau_{D^{0}}$ level~\cite{pajeroRecentAdvancesCharm2022}.

The CPTV analysis can be carried out based on the 
the two-hadron decays of $D^0$ mesons: $D^0\rightarrow K^-\pi^+$ and its CPT-conjugate $\overline{D^0}\rightarrow K^+\pi^-$.
The $D^0\rightarrow K^-\pi^+$ decay is dominated by the direct Cabibbo-favoured (CF) decays of $D^{0}$ mesons, which makes it almost fully flavour-specific and thus provides a clean probe for testing CPT symmetry.
The only contamination is due to the cases where $D^{0}$ mesons oscillate into $\overline{D^0}$ and subsequently decay via a Doubly-Cabibbo suppressed (DCS) mode $\overline{D^0}\rightarrow K^{-} \pi^{+} $. 
However, the smallness of $D^0\rightarrow K^{+} \pi^{-} $ branching fraction, $1.36 \times 10^{-4}$ for DCS as compared to $3.947 \times 10^{-2}$ ~\cite{particledatagroupReviewParticlePhysics2022}  for CF, combined with the slow mixing (oscillation period is three orders of magnitude larger than $D^0$ decay time) makes this admixture so small that it is usually neglected in the analyses. 

The central idea of the  CPT tests is the construction of a time-dependent decay-rate asymmetry. 
\begin{equation}
  A_{RS}(t) = \frac{N(D^0\to K^-\pi^+)(t)-N(\overline{D^0}\to K^+\pi^-)(t)}{N(D^0\to K^-\pi^+)(t)+N(\overline{D^0}\to K^+\pi^-)(t)}\label{eq:A_RS}.
\end{equation}
and extract the CPT-violating parameter $z$ by fitting it to the model, which for the flavour-specific channels can be expressed as~\cite{kosteleckyCPTLorentzViolation2001}::
\begin{align}
   &A_{\text{CPT}}(t) =\nonumber\\& ~~A_{\text{dir}} + \frac{2\Re(z)\sinh\Delta\Gamma t/2 - 2\Im(z)\sin\Delta mt}{(1+|z|^2)\text{cosh}\Delta\Gamma t/2 + (1-|z|^2)\text{cos}\Delta m t},
\label{eq:ACPT_fs}
\end{align}
where all asymmetries, including the direct CPTV term $A_{\text{dir}}$,
 are assumed to be much smaller than one. In this case, the measurement of the CPTV is independent of the CPV effects.

Due to the slowness of the $D^{0} \bar{D}^{0}$ oscillations, the asymmetry can be further simplify \eqref{eq:ACPT_fs} by approximating it to the leading order in ($x \tau \Gamma \ll 1$ and $y \tau \Gamma \ll1$) parameters : 

\begin{equation}
\label{eq:ACPTlinear}
  A_{\text{CPT}}(t) = A_{\text{dir}} + (\Re(z) y - \Im(z) x ) \Gamma t   , 
\end{equation}

\section{New CPT-symmetry violation (classical) limits with charm mesons}

\begin{figure}[tbh]
    \centering
    \begin{minipage}{0.48\textwidth}
        \centering
        \includegraphics[width=\textwidth]{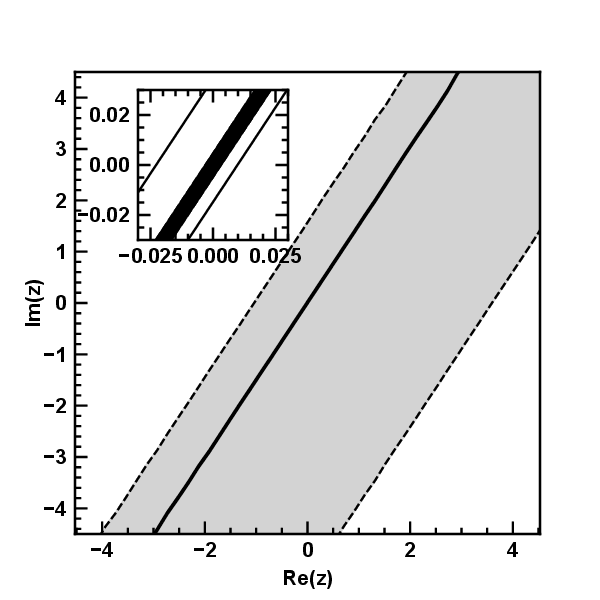}
    \end{minipage}
    \hfill
    \begin{minipage}{0.48\textwidth}
        \centering
        \includegraphics[width=\textwidth]{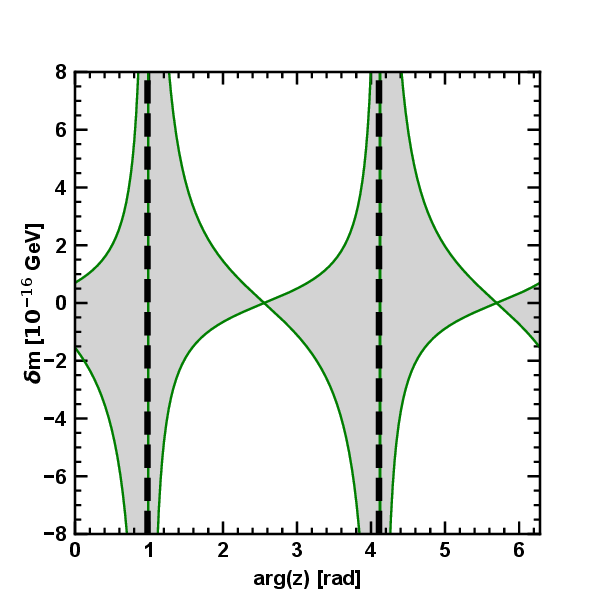}
    \end{minipage}
    \caption{\textbf{Left plot:} Experimental limits on the CPT-violating $z$ parameter at the 95\% confidence level (CL) in the $D^0 \to K^-\pi^+$ channel, as reported by the FOCUS Collaboration \cite{linkCharmSystemTests2003} (gray area) and inferred from the LHCb measurement \cite{aaijSearchTimedependentViolation2021a} (white area). The uncertainty arising from potential SM-compliant CPV is represented by the black area.  \textbf{Right plot:} Distribution of the mass difference $\delta m$ as a function of $\phi=\text{arg}(z)$. The gray area represents the 95\% CL, while the black dashed lines indicate the values of $\phi$ for which $\delta m$ is undefined due to the vanishing of the linear asymmetry term. The figures are adapted from ~\cite{krzemienNewConstraintsSymmetry2024}}
\label{fig:upper_limits} 
\end{figure}

Recently, new limits on  CPTV parameters in the charm sector have been established~\cite{krzemienNewConstraintsSymmetry2024}. The bounds were derived from the time-dependent asymmetry analysis of the $D^{0}\rightarrow K^{-}\pi^{+}$ decay channel, based on the data set collected by the LHCb collaboration~\cite{aaijSearchTimedependentViolation2021a}. The
slope of the linear fit can be identified as $y\Re(z) -x\Im(z)$. 
The result is statistically consistent with zero, indicating no evidence of CPTV  at the current sensitivity level.
The measured slope $s$ defines a linear relationship between the real and imaginary parts of the CPTV parameter z in the $(\Re(z), \Im(z))$ plane, given by:
$\Im{(z)} = y\Re{(z)}/x - s/x$ in the $(\Re(z), \Im(z))$ plane (see Fig.~\ref{fig:upper_limits}, left plot). 
This constraint can be further translated into limits on the mass and decay-width differences between flavour and antiflavour states (see Fig.~\ref{fig:upper_limits}, right plot). The resulting limits lie within the range $10^{-15} - 10^{-16}$ GeV, 
 representing the strongest classical constraint on CPTV in the charm sector to date.

At the present level of experimental precision, the dominant uncertainty arises from the limited knowledge of potential CP-violating effects. While their current impact remains subdominant, they could become significant in future analyses, particularly with increased statistics.

\section{Prospects of CPTV searches in charm with Standard Model Extension}
The Standard Model Extension~\cite{} is an established framework for CPTV searches across various physical systems.
It is built on the idea, originally provided by Greenberg~\cite{greenbergViolationImpliesViolation2002}, that any CPTV should imply Lorentz symmetry breaking. This concept was further developed by Kostelecky et al.~\cite{colladayCPTViolationStandard1997, kosteleckySensitivityCPTTests1998, kosteleckySignalsCPTLorentz1999} into an effective theory that systematically incorporates CPT- and Lorentz-violating terms,  while preserving the core properties of the "proper" quantum field theory. In the SME, the terms are organised as perturbative corrections to the known physics theory(Standard Model and General Relativity), making it both predictive and experimentally testable. 

This approach has been proven very fruitful for interpreting a wide range of experimental tests, as well as for proposing new searches, including those in the neutral flavour meson systems (\emph{c.f.}~\cite{datatables} for datatables).  

In the SME picture, the LV will manifest through the existence of a background tensor field filling the spacetime to which particles, e.g. mesons, can couple. This field plays the role of an "ether-like" structure, breaking isotropy and introducing dependencies on the particle boost and momenta directions. 
This, in turn, leads to the prediction of new experimental dependencies, such as sidereal variations, which arise due to the Earth's rotation, changing the experiment's orientation with respect to the fixed background field. These predicted dependencies not only increase the experimental sensitivity but also provide guidelines on how to look for the potential CPTV effect.

In particular, the CPTV parameter $z$ is not treated as a constant but depends on both the meson boost and momentum direction $\beta^{\mu}$, and can be expressed  as~\cite{kosteleckySignalsCPTLorentz1999}:

\begin{equation}
 z = \frac{\Delta a_{\mu}\beta^{\mu} }{\Gamma(x-iy)},
\end{equation}
where $\Delta a_{\mu}$ are real number coefficients.

The only experimental limits on CPTV in the charm sector to date were set by the FOCUS collaboration~\cite{linkCharmSystemTests2003}.
FOCUS analysed $D^{0}\rightarrow K^{-}\pi^{+}$ decay channel using data collected at FERMILAB from an experiment of high-energy photon beam with energy $\approx$~180 GeV incident on a fixed BeO target. The study was performed using both classical and SME-based approaches. The average Lorentz boost of the $D^{0}$ mesons in the sample was approximately equal to 39. The statistics were relatively modest, with around 35K events.

Between 2011 and 2018, the LHCb experiment collected $10^{4}$ times more events in the same decay mode. This dramatic increase in statistics, coupled with the similar Lorentz boost, suggests a potential improvement by a factor of approximately $100$ in terms of the sensitivity to the SME $\Delta a_{\mu}$ parameters for this channel.  

However, one can show that in the SME framework, the linear term $y\Re(z) -x\Im(z)$ accidentally vanishes due to the assumption that  $\Delta a_{\mu}$ are real.
Therefore, the asymmetry model becomes more complex;   higher-order terms must be taken into account. Also, previously neglected DCS decays become relevant,  and the CPV terms must be included.
An analysis of the LHCb dataset using the full SME-based model is currently ongoing.

\section{Conclusions and future perspective}
Neutral flavour meson oscillations are a fascinating quantum phenomenon that enables high-precision tests of fundamental symmetries, including CPT invariance. The neutral flavour meson systems offer experimental sensitivity that, in the case of kaons, is approaching the Planck scale.

Recently, we have established new limits on CPTV in the charm sector using the classical approach. Our results improve the previous FOCUS bounds by approximately two orders of magnitude. These bounds translate to the upper limits on the mass and decay width differences between flavour particle and antiparticle states of order $10^{-15}$ to $10^{-16}$ GeV at 95\% C.L.   

These results complement CPTV studies in other meson systems~\cite{particledatagroupReviewParticlePhysics2022}. 
The experimental precision to $\delta m$ and $\delta \Gamma$ is determined by the interplay of the relative values of the decay time scale and the mixing parameters, which differ considerably between meson families.
For instance, in the most favourable, neutral kaon system, the smallness of the denominator value in Eq.~(\ref{zpq})
 strongly amplifies sensitivity to $\delta m$ and $\delta \Gamma$ and reaches  a precision of about $10^{-18}$~GeV~\cite{angelopoulos1999}, while in the beauty sector, current bounds range from 
$\sim10^{-14}$~GeV \cite{Higuchi:2012kx, BaBar:2016zvy} for $B$, to  $\sim10^{-12}$~GeV for $B_s$~\cite{LHCb:2016vdl}. 
The CPT-violating coupling could, in principle, depend on quark content, making it valuable to explore all flavour sectors~\cite{kosteleckyCPTLorentzViolation2001}.


An ongoing analysis of LHCb data within the SME framework is further exploring CPT-violating effects in charm. 
It is worth mentioning that, due to recent theoretical progress, the data analyses using the non-minimal SME approach~\cite{edwardsSearchingCPTViolation2019} have also become feasible.

In the beauty sector, the data already recorded by LHCb is sufficient to improve previous CPTV limits within SME by at least a factor of two, with even greater potential as data taking at LHCb progresses.
The main limiting factor in CPT analyses in the charm sector remains the imprecisely known term describing potential CPV. However, this parameter may be measured with better precision in forthcoming experiments.
Among other ongoing experiments, 
the BESIII~\cite{BESIIIPublicWebpage, ablikimDesignConstructionBESIII2010} is currently operating at the $\psi(3770)$
resonance, has registered large numbers of quantum-correlated 
$D^{0} \bar{D}^{0}$ pairs. Unfortunately, since the resonance is produced at rest, without any boost, no analyses relying on lifetime measurements are feasible. 

The planned Super Tau-Charm Facility~\cite{Achasov:2023gey} is expected to significantly enhance sensitivity, combining high-statistics correlated charm production with boosted kinematics, which enables interferometric studies particularly well-suited for CPT studies.
Also, KLOE/KLOE-2~\cite{KLOE2PublicWebsite,amelino-cameliaPhysicsKLOE2Experiment2010}, which has recorded the world’s largest sample of entangled kaon pairs, is currently analysing its full dataset~\cite{antoniodidomenicoBackFutureQuantum2025,erykczerwinskiUpdatesCPSymmetry2025}, and might contribute to further CPTV searches.
Additionally, Belle II~\cite{BelleIIPublic,abeBelleIITechnical2010} is now taking data and might improve selected CPTV limits in the charm and beauty sectors, especially covering the channels with neutral particles in the final states. 

These datasets provide a rich and promising landscape to further tighten constraints and potentially discover the  CPT symmetry violation.

\bibliographystyle{unsrt}
\bibliography{CPT_indiana_2025_proceedings}


\section*{Acknowledgments}
W.K. wishes to thank the organisers for the kind invitation to present a talk and for the stimulating discussions during the meeting.  

\end{document}